\documentclass[a4,12pt]{article}
\usepackage{graphicx}
\newcommand{\nn}{\nonumber\\}
\newcommand{\bra}[1]{\left<#1 \right|}
\newcommand{\ket}[1]{\left| #1 \right>}

\newcommand{\Q}{Q_{\rm B}}

\renewcommand{\thepage}{}
\makeatletter
\@addtoreset{equation}{section}
\renewcommand{\theequation}{\thesection.\@arabic\c@equation}
\makeatother

\begin{document}
\begin{titlepage}
\title{
\vspace*{-4ex}
\hfill{\normalsize hep-th/0307173}\\
\vspace{4ex}
\bf Closed String Amplitudes
in Open String Field Theory\\
\vspace{5ex}}
\author{Tomohiko {\sc Takahashi}$^{1,}$\footnote{E-mail address:
 tomo@asuka.phys.nara-wu.ac.jp}\ \ 
and Syoji {\sc Zeze}$^{2,}$\footnote{E-mail address:
zeze@sci.osaka-cu.ac.jp}
\vspace{3ex}\\
$^1${\it Department of Physics, Nara Women's University, Japan}\\
$^2${\it Department of Physics, Osaka City University, Japan}}
\date{July, 2003}
\maketitle
\vspace{7ex}

\begin{abstract}
\normalsize
\baselineskip=19pt plus 0.2pt minus 0.1pt
We investigate
gauge invariant operators corresponding to on-shell closed string states
in open string field theory.
Using both oscillator representation and conformal mapping techniques,
we calculate a two closed string tachyon 
amplitude that connects two gauge
invariant operators by an open string propagator.
We find that
this amplitude is in a complete agreement with 
the usual disc amplitude.
\end{abstract}
\end{titlepage}

%%%%%%%%%%%%%%%%%%%%%%%%%%%%%%%%%%%%%%%%%%%%%%%%%%%%%%
\renewcommand{\thepage}{\arabic{page}}
\setcounter{page}{1}
\baselineskip=19pt plus 0.2pt minus 0.1pt
%%%%%%%%%%%%%%%%%%%%%%%%%%%%%%%%%%%%%%%%%%%%%%%%%%%%%%
%%%%%%%%%%%%%%%%%%%%%%%%%%%%%%%%%%%%%%%%%%%%%%%%%%%%%%
%%%%%%%%%%%%%%%%%%%%%%%%%%%%%%%%%%%%%%%%%%%%%%%%%%%%%%
\section{Introduction}

String field theory is a fascinating formulation to investigate
non-perturbative aspects of string theory. The progress in the last
three years in string field theory
\cite{rf:SenUniv,rf:Sen,rf:SZ-tachyon,rf:MT,rf:GR}, which has been
brought by calculational techniques of level truncation scheme
\cite{rf:KS-tachyon}, enables us to
describe the decay process of D-branes in terms of classical solutions
in the theory.

One of the most interesting facts in open string field theory is that,
as pointed out at the early stage formulating covariant string field
theories \cite{rf:HIKKO,rf:HIKKOloop,rf:CSFT,rf:GMW}, closed strings
are concealed in the theory as a bound state of the 
open string, because
open string loop diagrams look like closed string
propagations if we change time-slice.
Indeed we can see a tower of
closed string poles explicitly in non-planar one loop
amplitudes \cite{rf:FGST}.
While there are some recent works related to closed strings in open
string field theory \cite{rf:GIR,rf:EST}, the nature of closed strings
in the theory is not fully clear. 
In particular, we can not identify
closed strings with asymptotic states in open string field theory,
though such states are required by unitarity of the theory.
If we introduce on-shell closed strings as gauge invariant operators 
in open string field theory, the theory generates a single cover of
moduli space of Riemann surfaces with punctures and boundaries
\cite{rf:Zwiebach}. However, we can not produce purely closed string
amplitudes in the theory.

In order to describe purely closed string processes, an
interesting proposal
was made in the papers \cite{rf:Drkr1,rf:Drkr2} by Drukker.
If we obtain the tachyon vacuum solution and
the theory expanded around the solution,
we may reproduce closed string amplitudes by the Feynman
rules of the theory with the gauge invariant operators as sources. 
Because the propagator in the theory
around the solution does not generate any propagation of
boundaries of strings.
For this scenario, a possible candidate for the 
solution was constructed in Ref.~\cite{rf:TT} and some desirable results
were obtained in Refs.~\cite{rf:KT,rf:tomo,rf:TZ}.

In this paper we investigate the gauge invariant operators corresponding
to on-shell closed string states in order to develop techniques
for describing purely closed string processes.
Intuitively we expect that the operators can be constructed by locating
a closed string vertex operator at the midpoint on the identity
string field. There are, however, some subtleties that ghost operators
in the vertex operators diverge at the midpoint. To avoid this
difficulty, a regularization method is proposed in Ref.~\cite{rf:HI}.
We show that
this regularized gauge invariant operator 
agrees with the operator defined by the
method of conformal field theory.

In section 2, we define the gauge invariant operators making use of
conformal field theory.
We extend the operator for the closed string
tachyon given in Ref.~\cite{rf:HI} 
to that of D-branes background.
In section 3, we calculate a two closed string
tachyon amplitude using the Feynman rules and the gauge invariant
operator. Mapping the world sheet into a whole complex plane
explicitly, we show that the amplitude derived from the gauge invariant
operator agrees with the usual disc amplitude.
Moreover, we calculate the amplitude using the operator formalism
and we find that the result agrees with the same
amplitude. We give summary and discussion in section 4.

%%%%%%%%%%%%%%%%%%%%%%%%%%%%%%%%%%%%%%%%%%%%%%%%%%%%%%
%%%%%%%%%%%%%%%%%%%%%%%%%%%%%%%%%%%%%%%%%%%%%%%%%%%%%%
%%%%%%%%%%%%%%%%%%%%%%%%%%%%%%%%%%%%%%%%%%%%%%%%%%%%%%
\section{Closed String Source Terms}

Open string field theory with a midpoint interaction is defined by the
action \cite{rf:CSFT}
\begin{eqnarray}
 S_{\rm o}=\int \left(\Psi*\Q\Psi+\frac{2}{3}\Psi*\Psi*\Psi\right).
\end{eqnarray}
%The Feynman rules of this action reproduce open string amplitudes
%correctly. 
We can incorporate closed strings in the theory by adding
a source term to the action \cite{rf:Zwiebach,rf:GRSZ,rf:HI}:
\begin{eqnarray}
 S=S_{\rm o}+\int \Phi \Psi,
\end{eqnarray}
where $\Phi$ corresponds to an on-shell closed string state.
The Feynman rules derived from the action provide a single cover of
moduli space of open and closed strings amplitudes
\cite{rf:Zwiebach}.

Intuitively, we expect that this source term is represented by a
closed string vertex operator on the identity string field
\cite{rf:GRSZ,rf:HI}:
\begin{eqnarray}
 \int \Phi \Psi=\bra{I}V({\textstyle \frac{\pi}{2}})\ket{\Psi},
\end{eqnarray}
where $V(\sigma)$ is a closed string vertex operator
\begin{eqnarray}
 V(\sigma)=c_+(\sigma) c_-(\sigma) O(\sigma).
\end{eqnarray}
However, the midpoint ghost insertion has a singularity on the identity
and then we must regularize this expression to be  well-defined.
In Ref.~\cite{rf:HI}, it is proposed that a regulated source term should
be defined as
\begin{eqnarray}
\label{Eq:HIdef}
 \int \Phi \Psi
=\langle {\textstyle \tilde{I}}|\,O({\textstyle \frac{\pi}{2}})\ket{\Psi}, 
\end{eqnarray}
where $\tilde{I}$ denotes the state in which the anti-ghost
factors of the identity state are removed. In this definition,
we does not include the infinite factor arising
from the commutator of the ghost and anti-ghost fields.

If faithful to Ref.~\cite{rf:Zwiebach}, we should define
closed string sources in terms of conformal 
field theory. First we define the state  $\bra{V}$
for the closed vertex operator $V(\sigma)$ as
\begin{eqnarray}
\label{Eq:def}
\bra{V}\ket{A}=\langle\,h[V({\textstyle \frac{\pi}{2}})\,A(0)]\,\rangle,
\end{eqnarray}
where the state $\ket{A}$ is created by an operator as
$\ket{A}=A(w=0)\ket{0}$ and $h$ denotes the conformal transformation
induced by the function $h(w)=2w/(1-w^2)$, which maps a unit half
disc to an upper half plane \cite{rf:RZ}, and $\langle \cdots \rangle$
denotes correlation function on the upper half plane. Using this state,
we should define the closed string source term as
\begin{eqnarray}
\label{Eq:sourcedef}
 \int \Phi*\Psi=\bra{V}\ket{\Psi}.
\end{eqnarray}
By definition, it is easily seen that this source term
is a gauge invariant operator. Moreover,
it is known that
a correct amplitude is reproduced by level truncation analysis
based on this definition.\cite{rf:AG}

From (\ref{Eq:def}), the gauge invariant operator for 
a closed string tachyon is defined as
\begin{eqnarray}
\label{Eq:deftachyon}
 \bra{T;\,p,\,k}\ket{A}
=i^{-1}\,\langle\,c(i)\,c(-i)\,e^{ip\cdot X(i,-i)}\, 
e^{ik\cdot X(i,-i)}\,
h[A(w=0)]\,\rangle,
\end{eqnarray}
where the correlation function is defined on the whole complex plane
using the standard doubling trick.
Here, we consider string field theory on a D{\it p} brane and then we
impose Neumann boundary conditions to $X^\mu(z,\bar{z})\ 
(\mu=0,\cdots,p+1)$ and Dirichlet boundary conditions to 
$X^i(z,\bar{z})\ (i=p+1,\cdots,25)$. 
$p^\mu$ and $k^i$ denote the momenta in tangential and orthogonal
directions to a D{\it p} brane, respectively.

From the definition (\ref{Eq:deftachyon}), the oscillator expression of
the gauge invariant operators 
for a closed string tachyon is given by
\begin{eqnarray}
\label{Eq:tachyonstate}
 \bra{T;\,p,\,k}\ket{\Psi}&=&
\bra{-p}c_{-1}c_0\,\exp\left[(\alpha'p^2-1)\log4+
\sum_{n=1}^\infty(-1)^n\Big\{
-\frac{1}{2n}
\alpha_n\cdot\alpha_n-c_nb_n\right. \nn
&&
\hspace{1cm}\left.
-\frac{\sqrt{2\alpha'}}{n}p\cdot \alpha_{2n}
-\frac{2i\sqrt{2\alpha'}}{2n-1}k\cdot \alpha_{2n-1}
\Big\}\right]\ket{\Psi}.
\end{eqnarray}
This is a extended expression from that of Ref.~\cite{rf:HI}
to the case that contains Dirichlet directions.
The detail derivation of this expression will appear in a
forthcoming paper \cite{rf:TZ3}. Here, we only comment on the BRS
invariance of this state.
Though the BRS invariance is manifest by the definition
(\ref{Eq:deftachyon}), we can check it explicitly by the oscillator
expression. 
From (\ref{Eq:tachyonstate}), we find
\begin{eqnarray}
 K_{2N}^{\rm matt}\ket{T;\,p,\,k}=
4\alpha'(p^2+k^2)(-1)^N N\ket{T;\,p,\,k},
\end{eqnarray}
where the operator $K_{2N}^{\rm matt}$ is defined as 
$K_{2N}^{\rm matt}=L_{2N}^{\rm matt}-L_{-2N}^{\rm matt}$
using the matter Virasoro generators $L_n^{\rm matt}$. 
Similarly we obtain $K_{2N-1}^{\rm matt}\ket{T;\,p,\,k}=0$.
Therefore using the method in Refs.~\cite{rf:GJ,rf:HI}, we
conclude that, if we impose the on-shell condition $p^2+k^2=4/\alpha'$,
the state (\ref{Eq:tachyonstate}) is BRS invariant even for 
the case containing the Dirichlet directions.

%%%%%%%%%%%%%%%%%%%%%%%%%%%%%%%%%%%%%%%%%%%%%%%%%%%%%%
%%%%%%%%%%%%%%%%%%%%%%%%%%%%%%%%%%%%%%%%%%%%%%%%%%%%%%
%%%%%%%%%%%%%%%%%%%%%%%%%%%%%%%%%%%%%%%%%%%%%%%%%%%%%%
\section{Closed String Tachyon Amplitudes}

We consider the closed tachyon amplitude derived from the
source 
term (\ref{Eq:deftachyon}).
The simplest example is
derived from the diagram connecting two tachyon sources by an open
string propagator:
\begin{eqnarray}
\label{Eq:amp2}
(2\pi)^{p+1}\,\delta^{p+1}(p+p')\,
{\cal A}=\int_0^\infty dt\,
{}_1\langle\,T; \,p,\,k\,|
{}_2\langle\,T; \,p',\,k'\,|
\,b_0^{(1)}\,e^{-tL_0^{(1)}}\,|R\rangle_{12}.
\end{eqnarray}

\begin{figure}[b]
\centerline{\includegraphics[width=12cm]{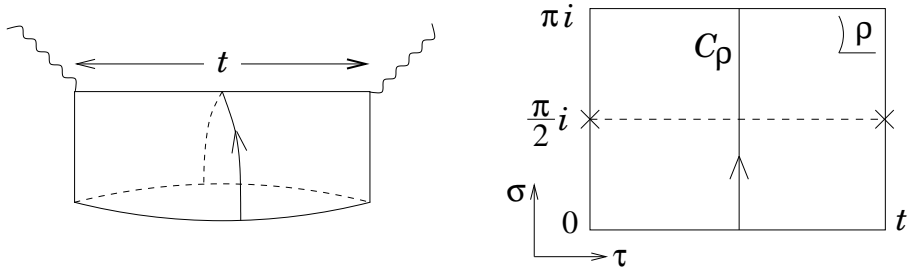}}
\caption{The world sheet corresponding to two closed tachyon amplitude.}
\label{fig:map1}
\end{figure}

Gluing the two surface states by an open string, we find the world sheet 
corresponding to the integrand in this amplitude
as depicted in Fig~\ref{fig:map1}. We introduce the coordinate
$\rho=\tau+i\sigma$ into this world sheet: the tachyon vertices insert
on $\rho=\pi\,i/2$ and $t+\pi\,i/2$, and the points $\sigma$ and
$\pi-\sigma$ are identified with each other at the boundaries
$\tau=0,\,t$. 
Consequently, we can rewrite the amplitude as
\begin{eqnarray}
\label{Eq:amp3}
{\cal A}&=&\int_0^\infty dt\,
\int_{C_\rho}\frac{d\rho}{2\pi i}\,
\langle\,b(\rho)\,c(\pi i/2)c(-\pi i/2)
c(t+\pi i/2)c(t-\pi i/2)\,\rangle_\rho \nn
&&
\times \langle\,
e^{ip\cdot X(\pi i/2,\,-\pi i/2)}
e^{ip'\cdot X(t+\pi i/2,\,t-\pi i/2)}\,\rangle_\rho 
\, \langle\,
e^{ik\cdot X(\pi i/2,\,-\pi i/2)}
e^{ik'\cdot X(t+\pi i/2,\,t-\pi i/2)}\,\rangle_\rho,
\end{eqnarray}
where we use the doubling trick and the contour $C_\rho$ is a path
from $\sigma=-\pi i$ to $\sigma=\pi i$ at an arbitrary $\tau$, and
$\langle\cdots\rangle_\rho$ denotes correlation function on the $\rho$
plane. 

\begin{figure}[t]
\centerline{\includegraphics[width=13cm]{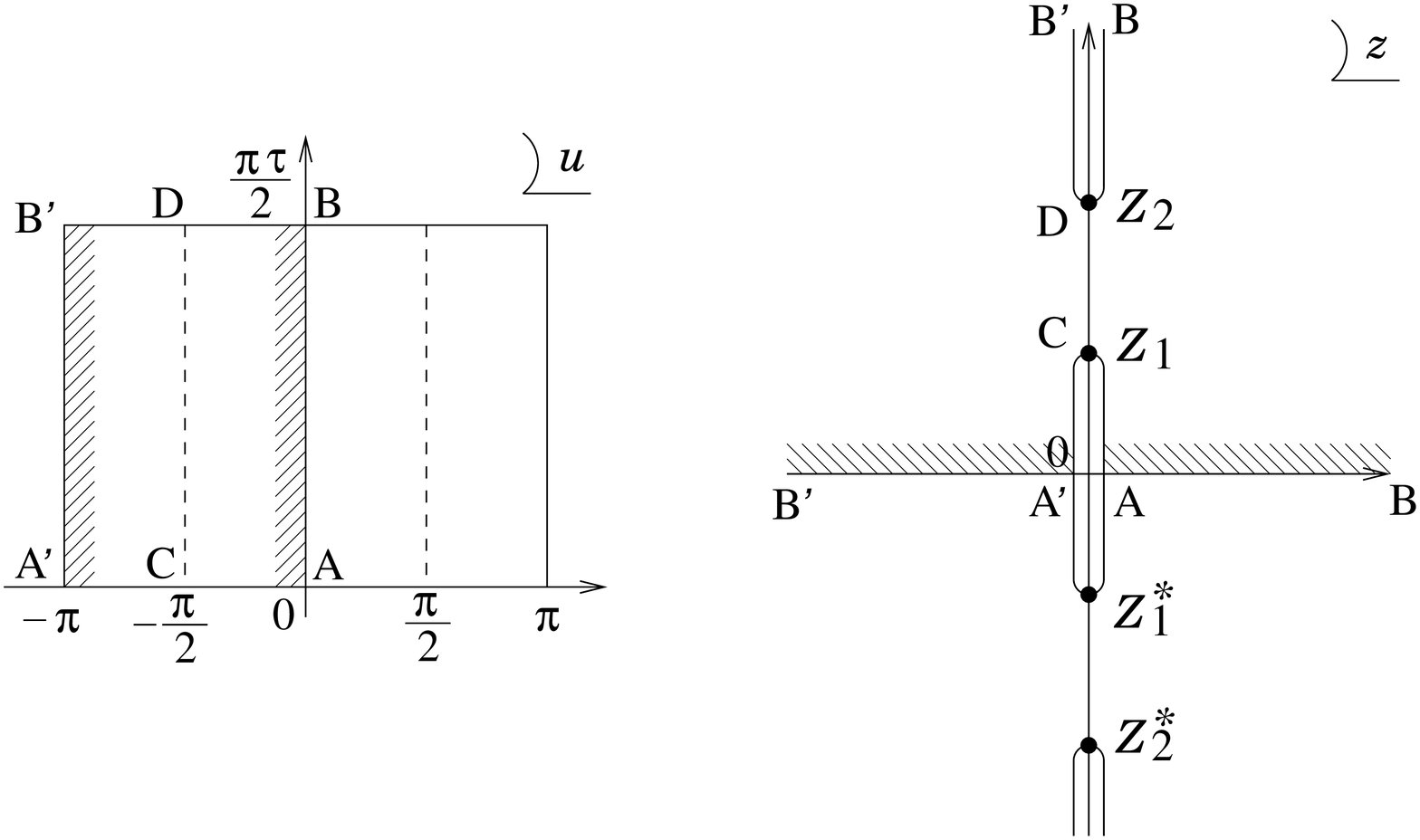}}
\caption{The $u$ and $z$ planes for the two closed tachyon amplitude.}
\label{fig:map2}
\end{figure}

Since the $\rho$ plane is topologically equivalent to a disc or after
the doubling trick to a sphere, it is more convenient to
map the $\rho$ plane to an upper half or a whole complex plane. First we
rotate the $\rho$ plane 90 degrees and put it on a region in the $u$
plane as depicted in Fig.~\ref{fig:map2}.
In the $u$ plane the left (right) region of the imaginary axis
corresponds to the holomorphic 
(anti-holomorphic) part of the open string, and the closed tachyon
vertices are inserted 
at the points C, D and their anti-holomorphic counter-points.
We introduce the parameter $\tau$ as $\pi \tau/2=it$.

We can transform this region in the $u$ plane to a whole complex $z$
plane by the mapping
\begin{eqnarray}
 z=-i\,\frac{\vartheta_1(u|\tau)}{\vartheta_4(u|\tau)},
\end{eqnarray}
where $\theta_i(u|\tau)\ \ (i=1,2,3,4)$ are theta functions with
periods $1$ and $\tau$ \cite{rf:WW}.
Using properties of theta functions, we find the relations
$z(-\pi-u)=z(u)$, $z(\pi-u)=z(u)$ and $z(0)=z(\pm\pi)=0$,
and also that $z$ is imaginary if $u$ is real. Therefore we find a cut
from $z(-\pi/2)$ to $z(\pi/2)$ in the $z$ plane and this cut corresponds
to the identification of AC and A$'$C in the $u$ plane.
The points $z(\pm \pi/2)$ are the insertion points of the closed tachyon
vertices in the $z$ plane and they are expressed as
\begin{eqnarray}
 z(-\pi/2)&=&-i\,
\frac{\vartheta_1(-\pi/2\,|\tau)}{\vartheta_4(-\pi/2\,|\tau)}
=i\,\frac{\vartheta_2}{\vartheta_3}=Z_1,
\nn
 z(\pi/2)&=&-i\,
\frac{\vartheta_1(\pi/2\,|\tau)}{\vartheta_4(\pi/2\,|\tau)}
=-i\,\frac{\vartheta_2}{\vartheta_3}=Z_1^*,
\end{eqnarray}
where $\vartheta_i$ denote $\vartheta_i(0|\tau)$  \cite{rf:WW}.
Moreover we can derive the relation $z(u+\pi\tau/2)=-1/z(u)$ and then we
find another cut from the points
$Z_2=i\vartheta_3/\vartheta_2$ and 
$Z_2^*=-i\vartheta_3/\vartheta_2$ to
the infinity along the imaginary axis. We can see also that the open
string boundary in the $u$ plane corresponds to the real axis in the $z$
plane.

The function $z(u)$ satisfies \cite{rf:WW}
\begin{eqnarray}
\label{Eq:difeq}
 \left(\frac{dz}{du}\right)^2=-(\vartheta_2^2+z^2\,\vartheta_3^2)
(\vartheta_3^2+z^2\,\vartheta_2^2).
\end{eqnarray}
This differential equation implies that the function $z(u)$ is an
elliptic function with the branch points $\pm
i\,\vartheta_2/\vartheta_3$ and $\pm i\,\vartheta_3/\vartheta_2$ in the
$z$ plane \cite{rf:WW}. Indeed the function $z(u)$ is an doubly-periodic
function with periods $2\pi$ and $\pi\tau$. Since the domain now
considered in the $u$ plane is half of the fundamental parallelogram,
it is clear that the domain is mapped to half domain of two sheets,
in this case the whole complex plane.
In addition, from the differential equation, it follows that
$dz/du \sim \sqrt{z-Z}$ near $z=Z=Z_i=Z_i^*$.
Then, if $z$ rotates by $2\pi$ around $Z$, $u$ rotates by $\pi$ around
its corresponding point, which is a conical singularity in the $u$ plane
and the vertex insertion point \cite{rf:GM}.
Thus we can map the domain in Fig.~\ref{fig:map2} to the complex plane.

\begin{figure}[b]
\centerline{\includegraphics[width=11cm]{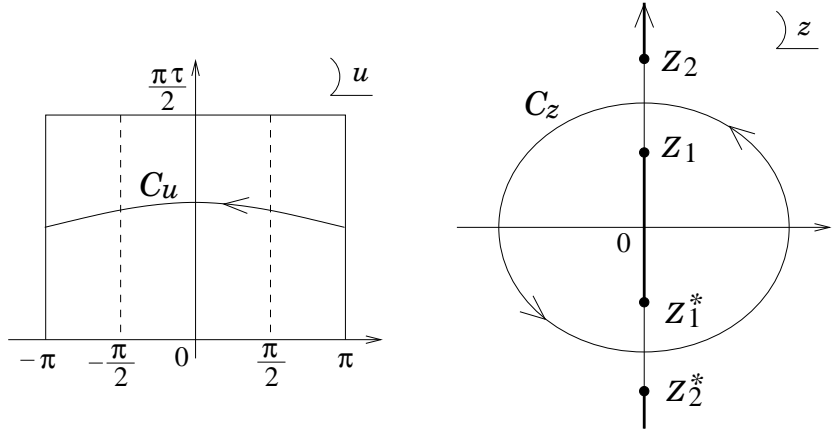}}
\caption{The contour $C_z$ in the equation (\ref{Eq:amp4})}
\label{fig:map3}
\end{figure}

Now that the conformal mapping is established, we can calculate the
amplitude in the $z$ plane. By the mapping $z(u)$ the amplitude
(\ref{Eq:amp3}) is transformed to
\begin{eqnarray}
\label{Eq:amp4}
 {\cal A}&=&\int_0^1 dx \left(\frac{dx}{dt}\right)^{-1}\,
\oint_{C_z} \frac{dz}{2\pi}\,\frac{dz}{du}\,\langle\,
b(z)c(Z_1)c(Z_1^*)c(Z_2)c(Z_2^*)\,\rangle
\nn
&&
\hspace{.5cm}
\times
\langle\,e^{ip\cdot X(Z_1,\,Z_1^*)}
\,e^{ip'\cdot X(Z_2,\,Z_2^*)}\,\rangle\,
\langle\,e^{ik\cdot X(Z_1,\,Z_1^*)}
\,e^{ik'\cdot X(Z_2,\,Z_2^*)}\,\rangle,
\end{eqnarray}
where conformal factors cancel using the on-shell condition,
and the contour
$C_z$ encircles the points $Z_1$ and $Z_1^*$ as depicted in
Fig.~\ref{fig:map3}. 
The modular parameter is given by $x=\vartheta_2/\vartheta_3$ and then
the vertex insertion points are written as $Z_1=ix$, $Z_2=i/x$,
$Z_1^*=-ix$ and $Z_2^*=-i/x$. As $t$ changes from $0$ to $\infty$, $x$
runs from $1$ to $0$.

The matter correlation function is calculated as
\begin{eqnarray}
\label{Eq:matamp}
&&
 \langle\,e^{ip\cdot X(Z_1,\,Z_1^*)}
e^{ip'\cdot X(Z_2,\,Z_2^*)}\,\rangle
=2^{-\alpha's}\left(
\frac{x^2}{1-x^4}
\right)^{-\alpha's},
\nn
&&
 \langle\,e^{ik\cdot X(Z_1,\,Z_1^*)}
e^{ik'\cdot X(Z_2,\,Z_2^*)}\,\rangle
=2^{-4-\alpha's}\left(
\frac{1-x^2}{1+x^2}
\right)^{-\alpha'(s+t/2)-4},
\end{eqnarray}
where the string coordinates of the Neumann directions are given by
$X^\mu(z,z')=(X^\mu(z)+\,X^\mu(\bar{z}))/2$
and that of the Dirichlet directions are given by
$X^i(z,z')=(X^i(z)-\,X^i(\bar{z}))/2$. The correlation
functions are calculated in terms of the propagators
\begin{eqnarray}
 \langle\, X^\mu(z)\,X^\nu(z') \,\rangle
&=&-2\alpha'\eta^{\mu\nu}\,\log(z-z'),\nn
 \langle\, X^i(z)\,X^j(z') \,\rangle
&=&-2\alpha'\delta^{i\,j}\,\log(z-z').
\end{eqnarray}
Here, we have defined kinematical invariants as $s=-p^2=-p'{}^2$ and
$t=-(k+k')^2$, and we have used the 
on-shell condition and the momentum
conservation $p^\mu+p'{}^\mu=0$.

The ghost correlation function is obtained as
\begin{eqnarray}
\langle\,
b(z)c(Z_1)c(Z_1^*)c(Z_2)c(Z_2^*)\,\rangle 
&=& -\frac{4(1-x^4)^2}{x^4(z^2+x^2)(z^2+x^{-2})}
\nn
&=&\,4\,\vartheta_3^4\,\frac{(1-x^4)^2}{x^2}\,
\left(\frac{dz}{du}\right)^{-2},
\end{eqnarray}
where we have used the differential equation (\ref{Eq:difeq}).
Using this result, we can perform the contour integration in
(\ref{Eq:amp4}): 
\begin{eqnarray}
\label{Eq:ghostamp}
\oint_{C_z} \frac{dz}{2\pi i}\,\frac{dz}{du}\,\langle\,
b(z)c(Z_1)c(Z_1^*)c(Z_2)c(Z_2^*)\,\rangle 
&=& 
\oint_{C_z} \frac{dz}{2\pi}\,\frac{dz}{du}\,
\,4\,\vartheta_3^4\,\frac{(1-x^4)^2}{x^2}\,
\left(\frac{dz}{du}\right)^{-2} \nn
&=&
\oint_{C_u} \frac{du}{2\pi}\,
\,4\,\vartheta_3^4\,\frac{(1-x^4)^2}{x^2}\nn
&=&
-4\,\vartheta_3^4\,\frac{(1-x^4)^2}{x^2}.
\end{eqnarray}

Next we evaluate $dx/dt$, which appears when we change the moduli
parameter from $t$ to $x$. The parameter $x=\vartheta_2/\vartheta_3$ is
given by $x=i\,z(\pi/2)$. Then we find
\begin{eqnarray}
\label{Eq:dxdtau}
 \frac{dx}{d\tau}
\left.
=\frac{\pi}{4}\,\frac{d^2z}{du^2}\,\right|_{u=\pi/2},
\end{eqnarray}
where we have used the equations
\begin{eqnarray}
 \frac{\partial \vartheta_i(u|\tau)}{\partial \tau}
=-\frac{i\pi}{4}
\frac{\partial^2 \vartheta_i(u|\tau)}{\partial u^2},
\end{eqnarray}
and $\vartheta_1'(\pi/2)=\vartheta_4'(\pi/2)=0$.\footnote{We denote
differentiations with respect to $u$ by primes.}
On the other hand,
differentiating both sides of Eq.~(\ref{Eq:difeq}) with respect to $u$,
we obtain
\begin{eqnarray}
 \frac{d^2 z}{du^2}=
-z(\vartheta_2^4+\vartheta_3^4+2\,\vartheta_2^2\,\vartheta_3^2\,z^2),
\end{eqnarray}
and, if $u=\pi/2$, we find
\begin{eqnarray}
\label{Eq:dzdu}
\left.
 \frac{d^2 z}{du^2}\right|_{u=\pi/2}
=i\,\vartheta_3^4\,x\,(1-x^4).
\end{eqnarray}
From Eqs.~(\ref{Eq:dxdtau}) and (\ref{Eq:dzdu}), it follows that
\begin{eqnarray}
\label{Eq:dxdt}
 \frac{dx}{dt}=-\frac{\vartheta_3^4}{2}\,x\,(1-x^4).
\end{eqnarray}

Putting Eqs.~(\ref{Eq:matamp}), (\ref{Eq:ghostamp}) and
(\ref{Eq:dxdt}) together, the amplitude is finally
obtained as 
\begin{eqnarray}
\label{Eq:amp5}
{\cal A}=\int_0^1dx\,
\frac{1-x^4}{2x^3}\,
\left\{\frac{4x^2}{(1+x^2)^2}\right\}^{-\alpha's}\,
\left(\frac{1-x^2}{1+x^2}\right)^{-\alpha't/2-4}.
\end{eqnarray}
We can rewrite this result to a familiar expression by a $SL(2,C)$
transformation. If we consider the transformation in the $z$ plane
\begin{eqnarray}
 z'=\frac{x^2-1}{x^2+1}\,\frac{z+i/x}{z-i/x},
\end{eqnarray}
the vertex insertion points $Z_i$ are transformed as $Z_1^*\rightarrow
\infty$, $Z_1\rightarrow 1$, $Z_2\rightarrow ((1-x^2)/(1+x^2))^2$ and 
$Z_2^*\rightarrow 0$. Then, 
writing $y=((1-x^2)/(1+x^2))^2$,
we find that the result agrees with the
amplitude of two closed tachyons on a disc \cite{rf:GIR,rf:AG}:
\begin{eqnarray}
 {\cal A}=\int_0^1dy\,y^{-\alpha't/4-2}\,
(1-y)^{-\alpha's-2}=B(-\alpha't/4-1,\,-\alpha's-1).
\end{eqnarray}

Let us calculate the scattering amplitude of two closed string
tachyons in terms of the oscillator expression (\ref{Eq:tachyonstate}).
The amplitude can be calculated as
\begin{eqnarray}
\label{Eq:amp6} 
{\cal A}
&=&
\int_0^\infty dt\,
4^{-2\alpha's-2}\,q^{-\alpha's-1}\,{\cal A}_{\rm gh}\,
{\cal A}_{\rm N}\,{\cal A}_{\rm D},\\
{\cal A}_{\rm gh}
&=&
\bra{0}c_{-1}\,\exp\left[-\sum_{n=1}^\infty
(-1)^nc_n b_n q^{2n}\right]\,
\exp\left[\sum_{n=1}^\infty(-1)^nc_{-n}b_{-n}\right]c_0c_1\ket{0}
\\
{\cal A}_{\rm N}
&=&
\bra{0}\,\exp\left[\sum_{n=1}^\infty
(-1)^n\left(
-\frac{1}{2n}{\alpha_{n}}_\mu \alpha_{n}^\mu\,q^{2n}
-\frac{\sqrt{2\alpha'}}{n}p_\mu\alpha_{2n}^\mu\,q^{2n}\right)
\right]\nn
&&
\hspace{1cm}
\times
\exp\left[\sum_{n=1}^\infty
(-1)^n\left(
-\frac{1}{2n}{\alpha_{-n}}_\mu\alpha_{-n}^\mu
+\frac{\sqrt{2\alpha'}}{n}p'_\mu\alpha_{-2n}^\mu\right)
\right]\ket{0}\\
{\cal A}_{\rm D}
&=&
\bra{0}\,\exp\left[\sum_{n=1}^\infty
(-1)^n\left(
-\frac{1}{2n}{\alpha_{n}}_i\,\alpha_{n}^i\,q^{2n}
-i\frac{2\sqrt{2\alpha'}}{2n-1}k_i\,
\alpha_{2n-1}^i\,q^{2n-1}\right)
\right]\nn
&&
\hspace{1cm}
\times
\exp\left[\sum_{n=1}^\infty
(-1)^n\left(
-\frac{1}{2n}{\alpha_{-n}}_i\,\alpha_{-n}^i
-i\frac{2\sqrt{2\alpha'}}{2n-1}k'_i\,
\alpha_{-2n+1}^i\right)
\right]\ket{0},
\end{eqnarray}
where we have introduced the parameter $q=\exp(-t)$, and
${\cal A}_{\rm gh}$, ${\cal A}_{\rm N}$ and ${\cal A}_{\rm D}$ denote
the contributions from ghost modes, the Neumann and Dirichlet direction
modes, respectively.
We can perform the contraction of oscillator modes in terms of a
familiar formula in 
Refs.~\cite{rf:HIKKO,rf:KP,rf:KK,rf:CG,rf:HT,rf:GS}.
The results are
\begin{eqnarray}
 {\cal A}_{\rm gh}&=&\prod_{n=1}^\infty(1-q^{2n}),
\\
{\cal A}_{\rm N}&=&\prod_{n=1}^\infty(1-q^{2n})^{-(p+1)/2}
\left[\prod_{n=1}^\infty
\frac{1-q^{4n}}{1-q^{4n-2}}\right]^{-4\alpha's},
\\
{\cal A}_{\rm D}&=&\prod_{n=1}^\infty(1-q^{2n})^{-(25-p)/2}
\left[\prod_{n=1}^\infty
\frac{1+q^{2n}}{1-q^{2n}}\right]^{-4\alpha's-16}
\left[\prod_{n=1}^\infty
\frac{1+q^{2n-1}}{1-q^{2n-1}}\right]^{4\alpha'(s+t/2)+16}.
\end{eqnarray}
Substituting these results into Eq.~(\ref{Eq:amp6}),
we obtain the two closed tachyon amplitude as
\begin{eqnarray}
\label{Eq:amp7}
 {\cal A}&=&
\int_0^\infty dt\,\frac{1}{16q}
\prod_{n=1}^\infty(1-q^{2n})^{-12}
\prod_{n=1}^\infty
\left\{
\frac{(1-q^{2n})}{(1+q^{2n})}
\right\}^{16}\nn
&&\times
\left[16q
\prod_{n=1}^\infty
\left\{\frac{(1-q^{4n})(1+q^{2n})(1-q^{2n-1})}
{(1-q^{4n-2})(1-q^{2n})(1+q^{2n-1})}\right\}^4
\right]^{-\alpha's}
\nn
&&
\times
\left[\prod_{n=1}^\infty
\frac{1-q^{2n-1}}{1+q^{2n-1}}\right]^{-2\alpha't-16}.
\end{eqnarray}

Finally, we show that two expressions (\ref{Eq:amp5}) and
(\ref{Eq:amp7}) of the amplitude are equivalent to each other.
To exhibit the dependence of $\vartheta_i$ on the
parameter $q$, we write it $\vartheta_i(q^2)$.
Using the formulas \cite{rf:WW}
\begin{eqnarray}
\label{Eq:koshiki1}
 \vartheta_3(q^{1/2})=\vartheta_3(q^2)+\vartheta_2(q^2),\ \  
 \vartheta_4(q^{1/2})=\vartheta_3(q^2)-\vartheta_2(q^2),\ \ 
 \vartheta_4(q^2)^4=\vartheta_3(q^2)^4-\vartheta_2(q^2)^4,
\end{eqnarray}
we find the equation
\begin{eqnarray}
\label{Eq:x2x2}
 \frac{1-x^2}{1+x^2}=\frac{\vartheta_3(q^{1/2})^2\,
\vartheta_4(q^{1/2})^2}{\vartheta_4(q^2)^4}.
\end{eqnarray}
We can express $\vartheta_i$ by infinite products;
\begin{eqnarray}
\label{Eq:koshiki2}
 \vartheta_3(q^{1/2})&=&G(q^{1/2})\prod_{n=1}^\infty
 (1+q^{(2n-1)/2})^2,\nn
 \vartheta_4(q^{1/2})&=&G(q^{1/2})\prod_{n=1}^\infty
 (1-q^{(2n-1)/2})^2,\nn
 \vartheta_4(q^2)&=&G(q^2)\prod_{n=1}^\infty
 (1-q^{4n-2})^2,
\end{eqnarray}
where $G(q)=\prod(1-q^{2n})$. Substituting these expressions into
Eq.~(\ref{Eq:x2x2}), we obtain the equation
\begin{eqnarray}
\label{Eq:xq1}
 \frac{1-x^2}{1+x^2}=\left[\prod_{n=1}^\infty
\frac{1-q^{2n-1}}{1+q^{2n-1}}\right]^4.
\end{eqnarray}
Similarly, using formulas related to the theta functions, we can find the
following equations
\begin{eqnarray}
\label{Eq:xq2}
&&
 -\frac{dx}{dt}\,\frac{1-x^4}{2x^3}
=\frac{1}{16q}\prod_{n=1}^\infty(1-q^{2n})^{-12}
\prod_{n=1}^\infty\left\{
\frac{1-q^{2n}}{1+q^{2n}}\right\}^{16},\\
\label{Eq:xq3}
&&
\frac{4x^2}{(1+x^2)^2}=
16q\left\{
\prod_{n=1}^\infty
\frac{(1-q^{4n})(1+q^{2n})(1-q^{2n-1})}{
(1-q^{4n-2})(1-q^{2n})(1+q^{2n-1})}
\right\}^4.
\end{eqnarray}
From Eqs.~(\ref{Eq:xq1}), (\ref{Eq:xq2}) and (\ref{Eq:xq3}), 
we can see that the amplitude (\ref{Eq:amp5}) derived from the method of
conformal field theory completely agrees with the expression
(\ref{Eq:amp7}) obtained by the oscillator representation.

\section{Summary and Discussion}

We have studied the gauge invariant operators in open string field
theory. We have defined the gauge invariant operators in terms of
conformal field theory and have shown the oscillator expression of
the operator corresponding to the on-shell closed string tachyon
state. Using this operator, we have calculated the two closed string
tachyon amplitude on a disc in terms of conformal techniques. The
resulting amplitude agrees with the usual disc amplitude. 
Moreover, we have calculated the amplitude in terms of the oscillator
representation of the operator. We have found that both derivations
lead us to the exactly same amplitude.
This coincidence indicates that the gauge invariant operator has
a well-defined oscillator expression,
although some subtleties often arise in the identity string field.

The gauge invariant operators discussed in this paper are also
observables in the theory expanded around universal solutions in
Ref.~\cite{rf:TT}, as pointed out by Drukker \cite{rf:Drkr2}. 
The propagator in the theory around a non-trivial universal solution is
given by $L=L_0'/2-(L'_2+L'_{-2})/4+3/2$, where $L'_n$ denote twisted
ghost Virasoro generators \cite{rf:TZ}. We can see that this propagator
generates closed surfaces and then we can regard it 
a kind of closed string propagators \cite{rf:Drkr2}. Therefore we could
derive the 
two closed string tachyon amplitude on the sphere from the
`string diagram' connecting the gauge invariant operators by the
propagator in the theory expanded around the universal solution.

We expect that the non-trivial universal solution can be identified with 
the tachyon vacuum solution. Because we have already proved the no open
string theorem which states that there is no open string excitation
perturbatively in the theory expanded around the solution
\cite{rf:KT,rf:TZ}. Another reason is that there is no
stable non-perturbative vacuum in the theory around the solution and this
fact implies that the theory is on the tachyon vacuum from the outset
\cite{rf:tomo}. In addition, the universal solution can be expressed by
a state in the universal Fock space \cite{rf:TT}. 
These facts are required for the tachyon vacuum solution. If we can
reproduce purely closed string amplitudes in the theory expanded around
the universal solution, we have further evidence for the correspondence
between the tachyon vacuum and the universal solution. 

\section*{Acknowledgements}
We would like to thank Hiroshi Itoyama for
encouragement.

%%%%%%%%%%%%%%%%%%%%%%%%%%%%%%%%%%%%%%%%%%%%%%%%%%%%%%%%%%%%%

\end{document}